\documentclass[twocolumn,prl,amsmath,amssymb,showpacs,superscriptaddress,floatfix]{revtex4}
\usepackage{graphicx}
\usepackage{bm}

\usepackage{epsf}
\begin{document}
\title{Possible phase transition in liquid Cesium at ambient pressure}

\author{Yu. D. Fomin, E. N. Tsiok, V. N. Ryzhov}
\affiliation{Institute for High Pressure Physics RAS, 108840
Kaluzhskoe shosse, 14, Troitsk, Moscow, Russia \\ }

\date{\today}

\begin{abstract}
We report a molecular dynamics study of liquid cesium at ambient
pressure intended to check the possibility of liquid-liquid phase
transformation at $T=590$ K. We find the presence of small kinks
on thermodynamic characteristics of the system, but no phase
transition.
\end{abstract}


\maketitle

\section{Introduction}

Cesium is an element which demonstrates extremely complex
behavior. The phase diagram of cesium contains several structural
transitions in crystalline region \cite{tonkov}. At low pressures
it forms an bcc phase (Cs-I) which transforms into fcc (Cs-II)
under pressure. Further pressurizing induces a transition into
Cs-III phase which has very narrow region of stability. More
structural transitions take place in Cs if the pressure is
increased further.

Interestingly, both Cs-I and Cs-II phases demonstrate maxima at
the melting line. It is widely believed that maxima on the melting
line evidence of the complex behavior of the liquid phase.
Moreover, from the common sense one expects that close to the
melting line the local structure of liquids should be similar to
the corresponding crystal. That is why it is possible that
structural transformations in the solid part of the phase diagram
could induce also transformations in the liquid.

Such changes in the properties of the liquid Cs were discovered
several decades ago. For example, in Ref. \cite{exp-1990} the
structure of liquid cesium at several pressures and temperatures
slightly above the melting line maxima was measured by X-ray
diffraction. The authors discovered even two transformations. The
first one at the pressure $P_1 \approx 2$ GPa corresponds to
changes from bcc-like liquid to the fcc-like one. The second
transition occurs at $P_2 \approx 3$ GPa. This transition
transforms the fcc-like liquid into even denser one. Further
evidence of the liquid-liquid phase transition (LLPT) in cesium
was reported in \cite{falconi}. X-ray measurements of structure of
liquid cesium at high pressure and high temperature were performed
and it was found that the density demonstrated a jump at $P=3.9$
GPa. However, in a recent work \cite{hattori} it is claimed that
the procedure for measuring the density in Ref. \cite{falconi} is
not enough precise and more accurate determination of the density
demonstrates no discontinuity.

Phase transitions in liquid cesium under pressure were studied in
a large body of publications. However, all these publications were
concerned to the transitions mentioned above. At the same time
several years ago several publications reported the possibility of
another phase transition in liquid cesium
\cite{blag-97,blag-2000,blag-2008,blag-2012}. The authors of these
publications performed measurements of adiabatic thermal pressure
coefficient of liquid cesium at the ambient pressure and the
temperatures up to 600 K. The adiabatic thermal pressure
coefficient is defined as:

\begin{equation}
   \gamma= \frac{1}{T} \left( \frac{\partial T}{\partial P}
   \right)_S.
\end{equation}
From thermodynamic relations it can be shown that $\gamma=
\frac{\alpha_P}{\rho c_P}$, where $\alpha_P$ is the thermal
expansion coefficient and $c_P$ is the isobaric heat capacity. A
small jump of $\gamma$ at the temperature $T_0=590$ K was observed
in \cite{blag-97,blag-2000,blag-2008,blag-2012}. Although this
jump is just about $2 \%$, the authors claim that it is within the
experimental uncertainties and make a conclusion that a phase
transformation takes place.

An important point about this transition is that it takes place at
ambient pressure and relatively low temperatures. Typically LLPT
in metals are related to electronic transformations. This
explanation was also used for explanation of possible LLPT in Ce
at high pressure. It is not clear whether electronic
transformation can take place at ambient pressure and temperature
about 600 K. However, LLPT can appear even in simple systems of
particles interacting via isotropic pair potentials (see, for
instance, \cite{boat,genvdw,skib,skib1,rs-pre,rs-jetp}). In this
case LLPT is induced by existence of two characteristic distances
in the interaction potential. Therefore, even at the conditions of
experiments from Refs.
\cite{blag-97,blag-2000,blag-2008,blag-2012} LLPT appears as a
transition from BCC-like liquid to FCC-like one. However, up to
now only one group reported observation of this phase transition
at $T=590$ K and independent measurements would be able to shed
light on the problem. There is also a theoretical work
\cite{ghatee} which pretends to describe the transition in liquid
Cs at ambient pressure by employing Lennard-Jones potential with
parameters which depend on the temperature. However, this model
was constructed specially to describe the results of experiments
on Refs. \cite{blag-97,blag-2000,blag-2008,blag-2012} and because
of this it cannot be considered as independent verification of the
results of these experiments.

In this paper we for the first time perform a simulational study
of the possible phase transition in liquid cesium at $T=590$ K and
ambient pressure in order to carry out an independent exploration
of the phenomenon by a different approach.

\section{System and Methods}

In the present study we simulated a system of 3456 particles in a
cubic box with periodic boundary conditions by means of molecular
dynamics method. The interaction of the particles is described by
embedded atom potential proposed by Belashchenko in Ref.
\cite{eam-cs}. This potential was specially designed for molecular
simulation of liquid cesium. In particular, in Ref. \cite{eam-cs}
a possibility of LLPT at $3.9$ GPa was questioned and no
indications of the transformation were observed. The initial
configuration of the particles was taken to be bcc lattice. We
studied the temperatures from $T_{min}=300$ K up to $T_{max}=1200$
K. At all these temperatures the system rapidly melts. The system
was simulated at constant pressure (isothermal-isobaric ensemble).
The pressure was set to 1 bar. Additional simulations at 100 and
1000 bars were also performed. The time step was set to 0.2 fs.
The equilibration run took $5 \cdot 10^6$ steps and the production
run - $10 \cdot 10^6$ steps. We calculated the equations of state
of the system and the internal energies. The response functions
(isobaric heat capacity $c_P$ and thermal expansion coefficient
$\alpha_P$) were obtained by numerical differentiation of these
functions. We also performed the calculations of structural
properties (radial distribution functions, structure factors and
angle distribution functions) and diffusion coefficients. The
later were computed via the Einstein relation.

\section{Results and Discussion}

\begin{figure} \label{cmp}
\includegraphics[width=8cm]{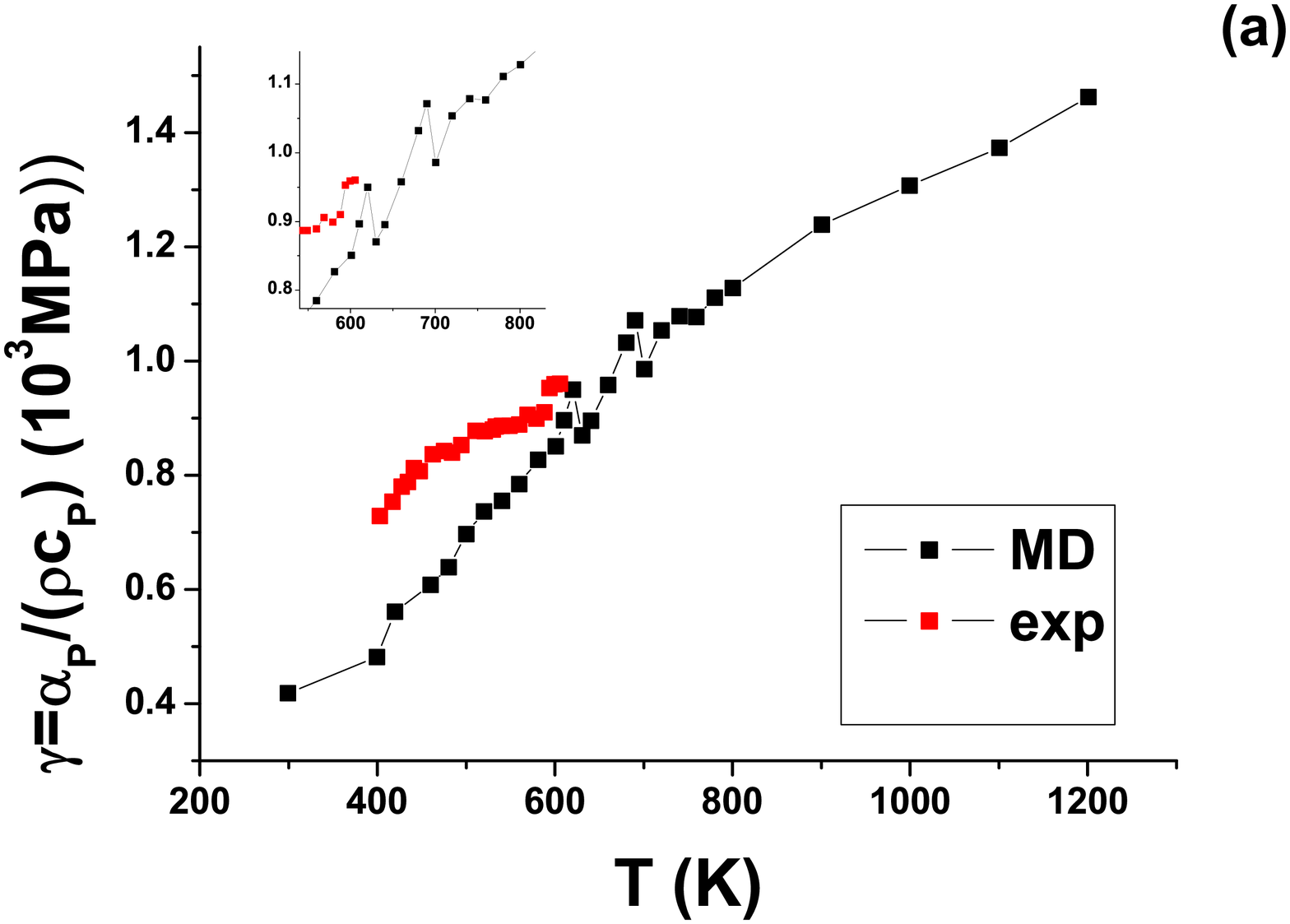}%

\includegraphics[width=8cm]{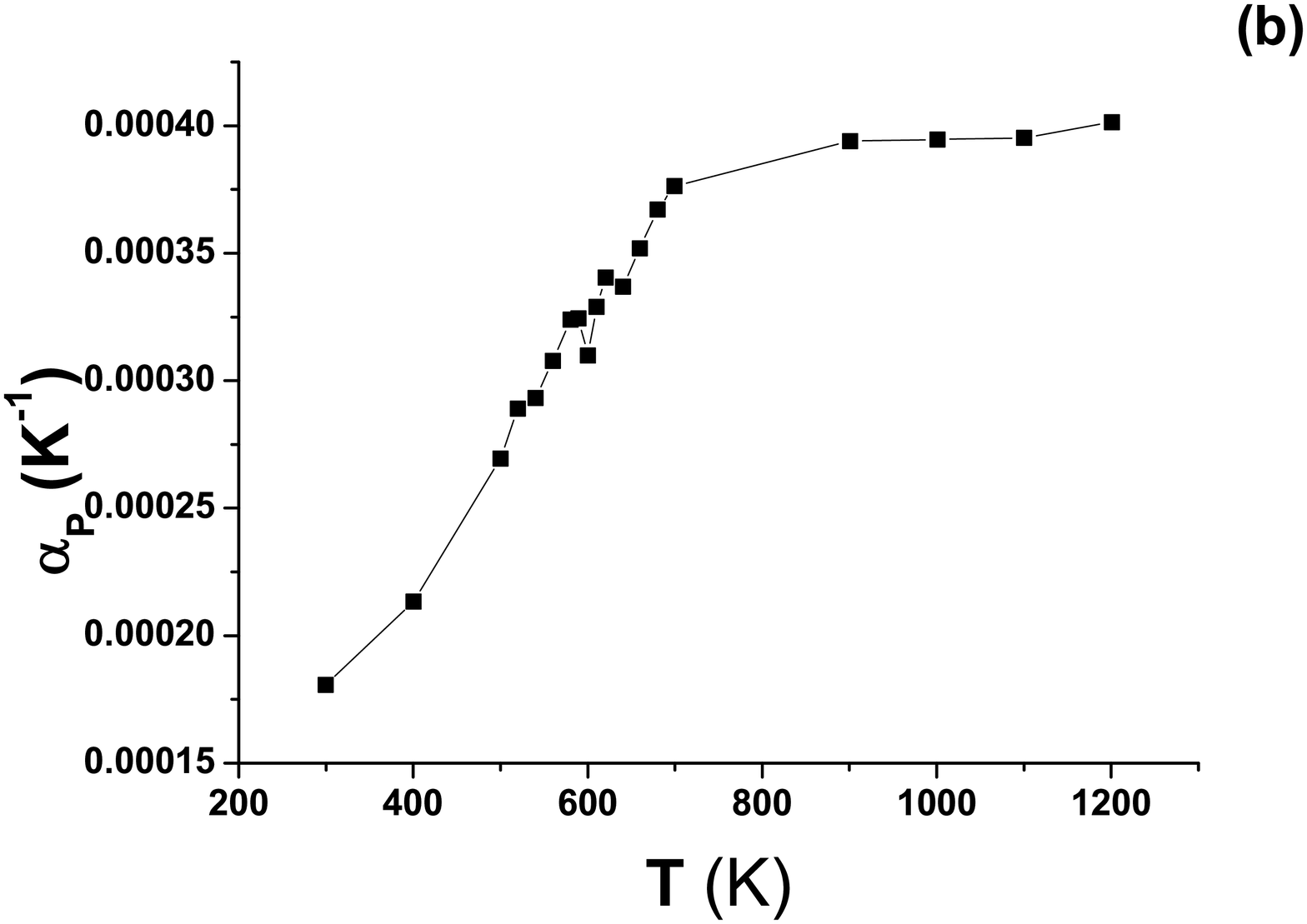}%

\includegraphics[width=8cm]{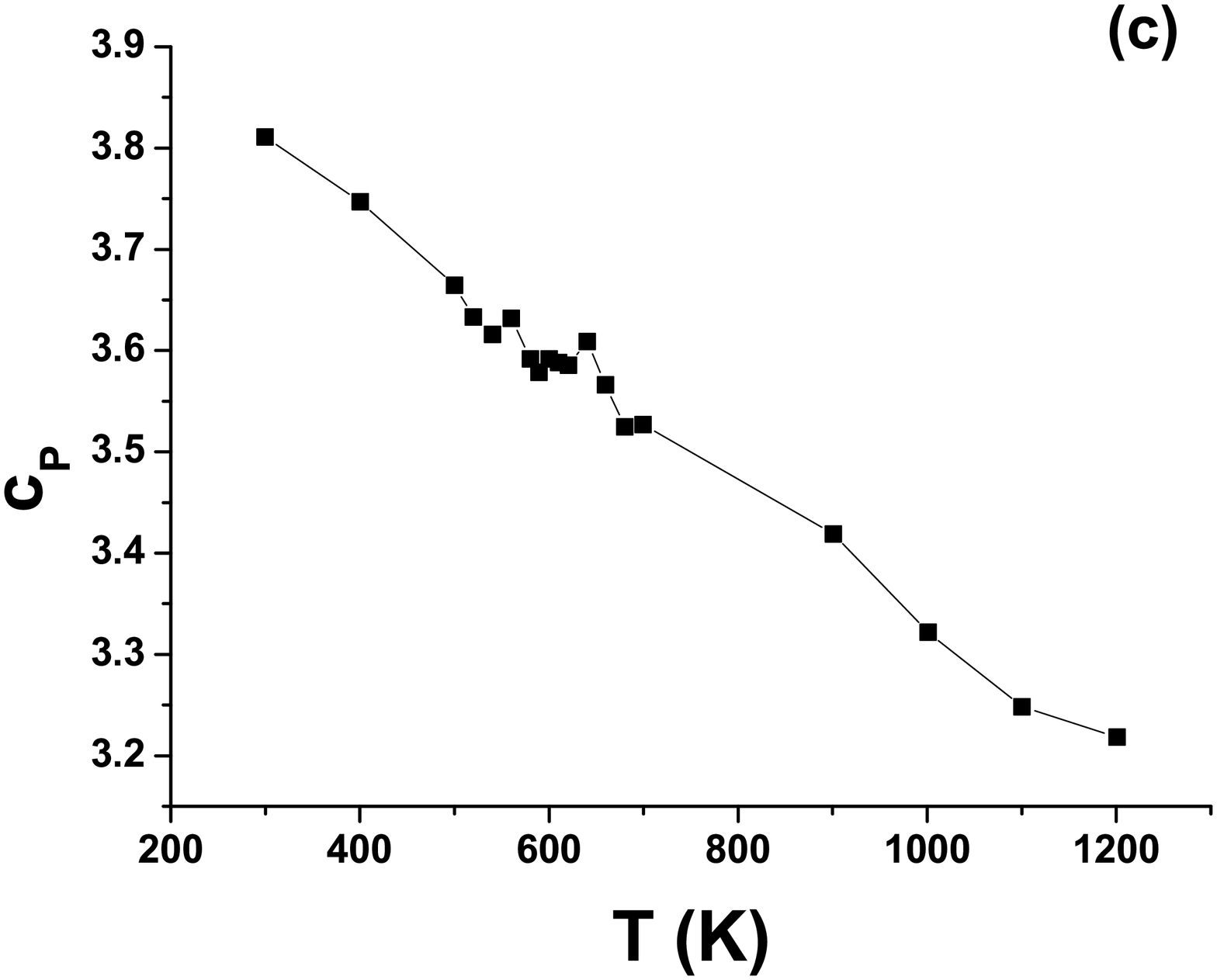}%

\caption{(a) Adiabatic thermal coefficient, (b) thermal expansion
coefficient and (c) isobaric heat capacity of liquid cesium at
$P=1$ bar. Experimental results from Ref. \cite{blag-2000} are
shown for comparison at panel (a).}
\end{figure}

First of all we calculate the equation of state along an isobar
$P=1.0$ bar from $T_{min}=300$ K up to $T_{max}=1200$ K. We also
performed calculations of enthalpy along this isobar and computed
the isobaric heat capacity $c_P$ by numerical differentiation. The
coefficient $\gamma$ was obtained as $\gamma=\frac{\alpha_P}{\rho
c_P}$, where the thermal expansion coefficient $\alpha_P$ was
obtained by numerical differentiation of the density $\alpha_P= -
\frac{1}{\rho} \left ( \frac{\partial \rho}{\partial T} \right
)_P$. The results for $\alpha_P$, $c_P$ and $\gamma$ are given in
Figs. ~\ref{cmp} (a)-(c). One can see that in spite of rather good
averaging the data are still noisy. The thermal expansion
coefficient demonstrates two kinks at $T=620$ and $T=690$ K (Figs.
~\ref{cmp} (b)). Also a small maximum and minimum are observed at
$T=900$ and $1100$ K respectfully. The isobaric heat capacity
looks smoother although it is also noisy in the interval $580-720$
K (Fig. ~\ref{cmp} (c)).

The behavior of the coefficient $\gamma$ is mostly governed by the
behavior of the thermal expansion coefficient (Fig. ~\ref{cmp}
(b)). It also demonstrates two small peaks at $T=620$ and $T=690$
K which are slightly above the assumed transition point.

A comparison with experimental values of $\gamma$ is also given in
Fig. ~\ref{cmp} (a). One can see that the data from simulations
are systematically lower. However, the kinks at the $\gamma$ from
simulations almost coincide with the experimental point ($T=620$ K
from simulation vs. $T=590$ K from experiment) which allows us to
assume that the simulation shows qualitative agrement with
experiment.

From Fig. ~\ref{cmp} (a) we cannot make a clear conclusion about
the presence of transition. Although we observe some kinks in the
adiabatic thermal coefficient they are just a bit above the range
of the errors of calculations. However, the effect observed in
experiments is also hardly visible. In order to see whether there
is a phase transition in the system we study its properties on
both sides of the possible transition.


\begin{figure} \label{eos-isot}
\includegraphics[width=8cm]{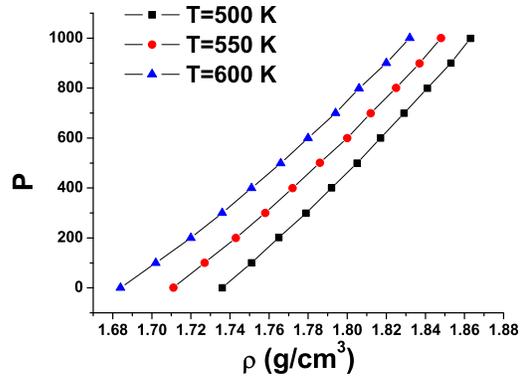}%

\caption{\label{eos} Isotherms of liquid cesium. From left to
right: $T=600$ K, $550$ K and $500$ K}.
\end{figure}

We study the equations of state along isotherms (Fig.
\ref{eos-isot}) and we do not observe anything which can be
interpreted as a phase transition.

\begin{figure}\label{gr-sk}
\includegraphics[width=8cm]{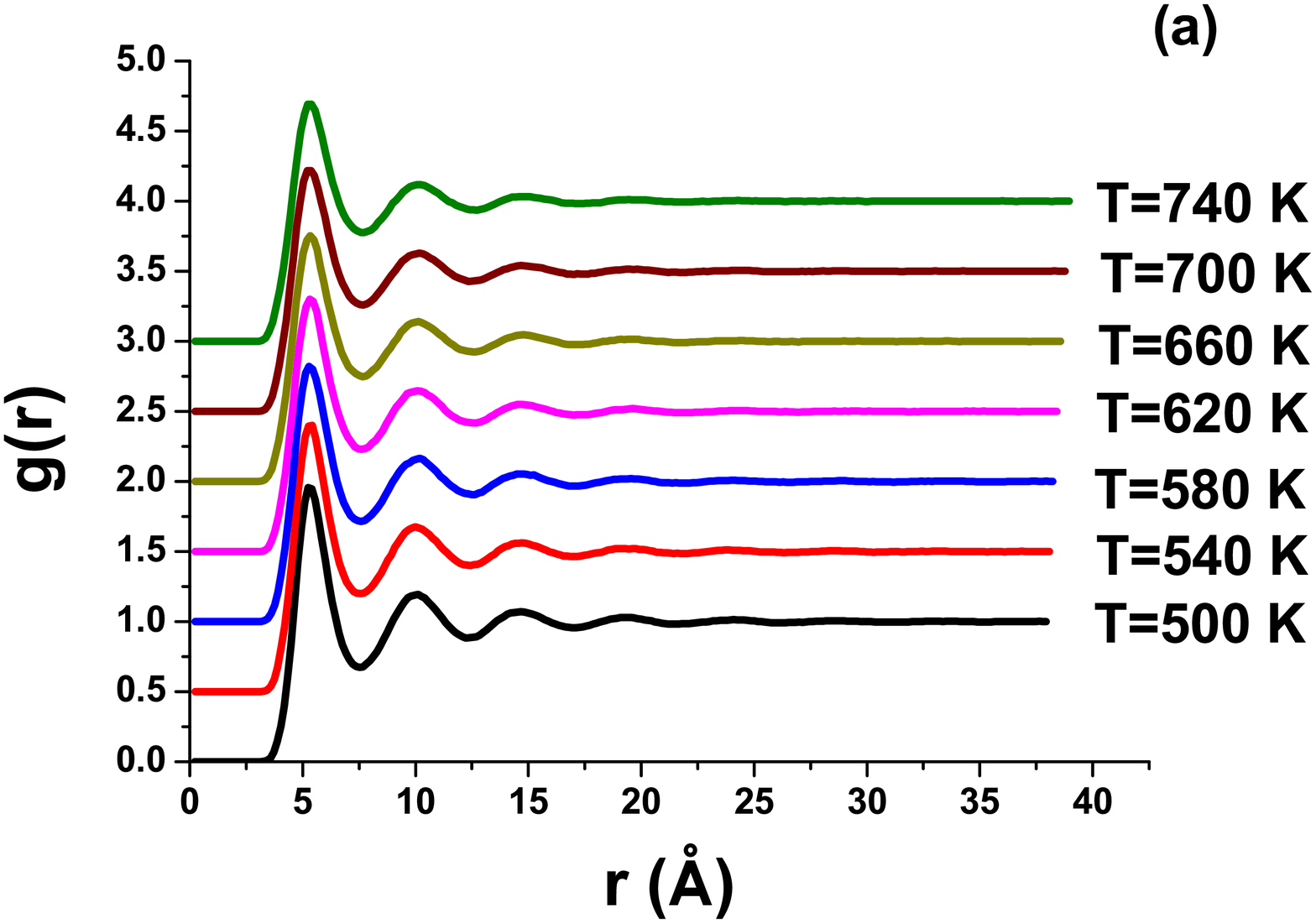}%

\includegraphics[width=8cm]{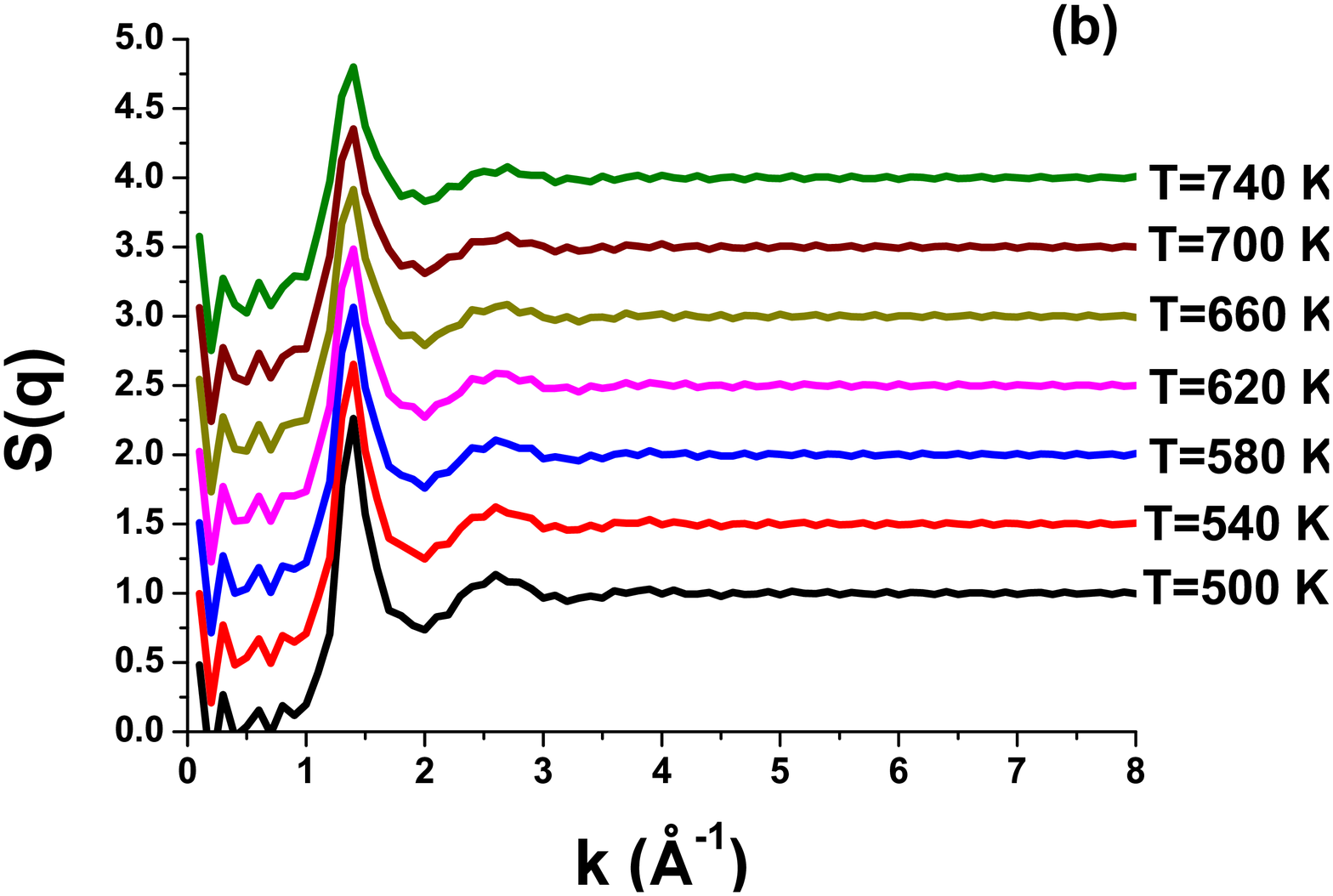}%

\includegraphics[width=8cm]{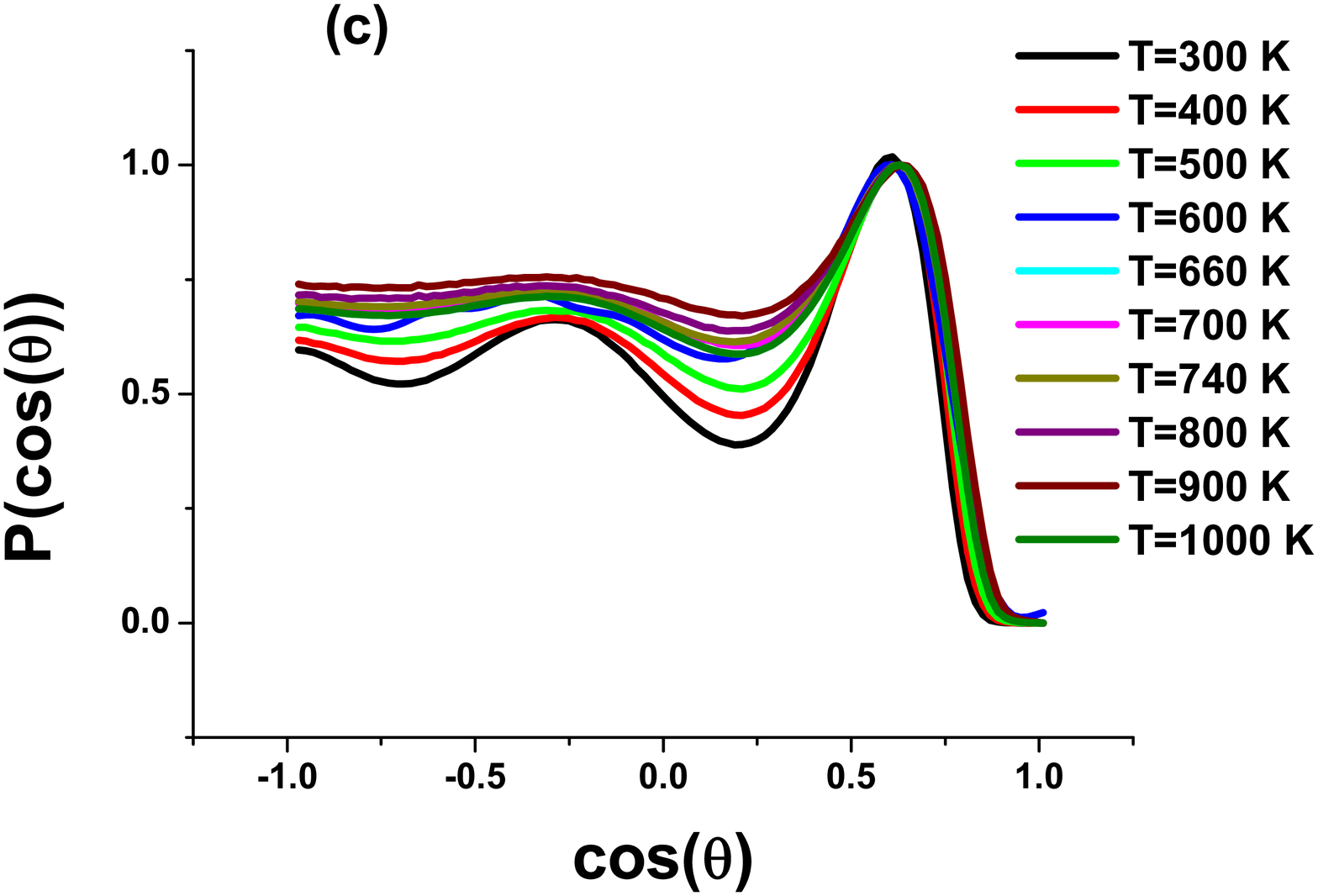}%

\caption{\label{sk} (a) Radial distribution functions; (b)
structure factors and (c) angle distribution function of liquid
cesium for a set of temperatures at $P=1$ bar isobar.}
\end{figure}

As the next step we study radial distribution functions and
structure factors of the system (Fig. \ref{gr-sk} (a) and (b)) and
we again do not observe any significant changes in the structure.
Some less pronounced structural changes can be observed from the
triplet correlation function which is related to the angle
distribution function, shown in Fig. \ref{gr-sk} (c). However,
this function also does not demonstrate any changes by passing the
hypothetical transition point.

\begin{figure}\label{nn-cs}
\includegraphics[width=8cm]{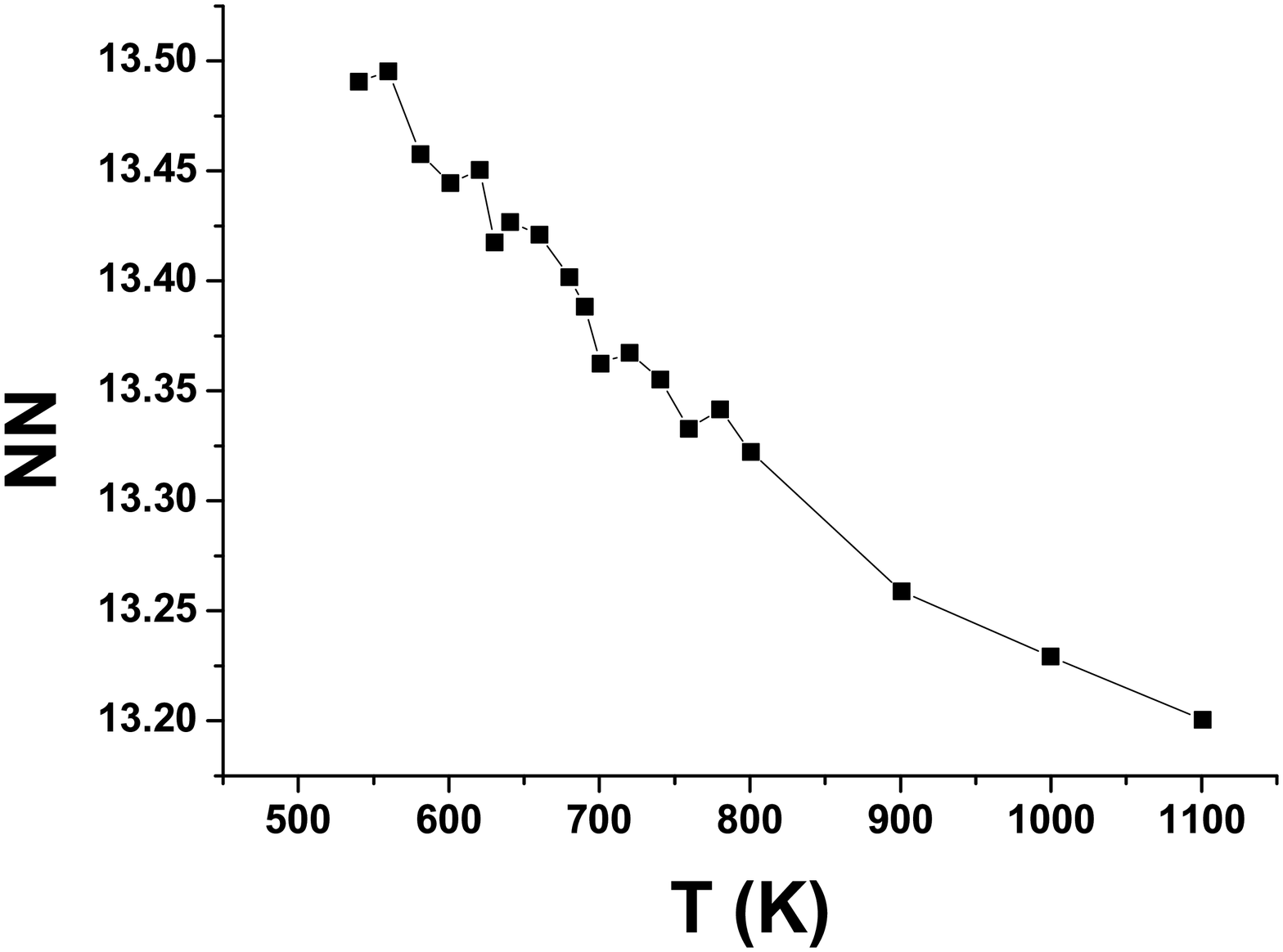}%

\caption{\label{sk} Number of nearest neighbors at $P=1$ bar
isobar.}
\end{figure}

Phase transformations are usually accompanied by jumps of the
nearest neighbor number (NN). The location of the first peak of
radial distribution function also can experience a jump. The
coordination number can be obtained as $NN=4 \pi \rho_N
\int_{0}^{r_m} r^2g(r)dr$, where $\rho_N=N/V$ is the number
density of the system and $r_m$ is the location of the first
minimum of $g(r)$. Fig. \ref{nn-cs} shows the number of nearest
neighbors at $P=1$ bar isobar. One can see that the NN is rather
noisy. One can find some kinks next to the kinks of $\gamma$,
however, they are within the accuracy of calculations. The
position of the first peak of $g(r)$ changes smoothly with
temperature.

\begin{figure}\label{diff-coef}
\includegraphics[width=8cm]{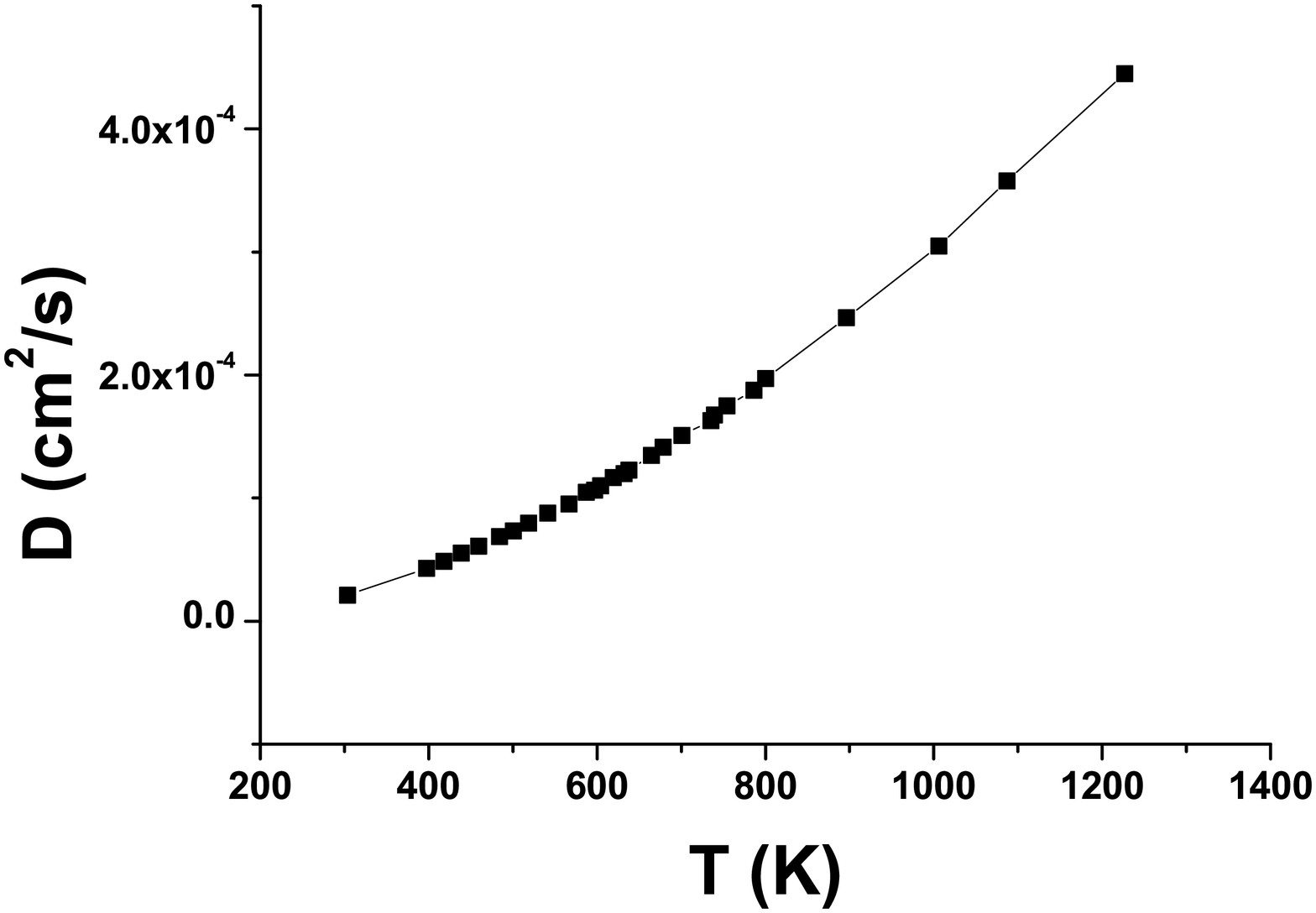}%

\caption{\label{diff} Diffusion coefficient of liquid cesium along
$P=1$ bar isobar.}
\end{figure}

We apply one more criterion which can establish the presence of
the phase transforation in the case of phase transition one can
expect that the mobility of particles can change which would lead
to the changes in diffusion coefficient. The diffusion coefficient
as a function of temperature along $P=1$ bar isobar is shown in
Fig. \ref{diff-coef}. Again we do not observe any changes in its
behavior. The temperature dependence of diffusion is smooth and
does not indicate any kind of phase transition.

\begin{figure}
\includegraphics[width=8cm]{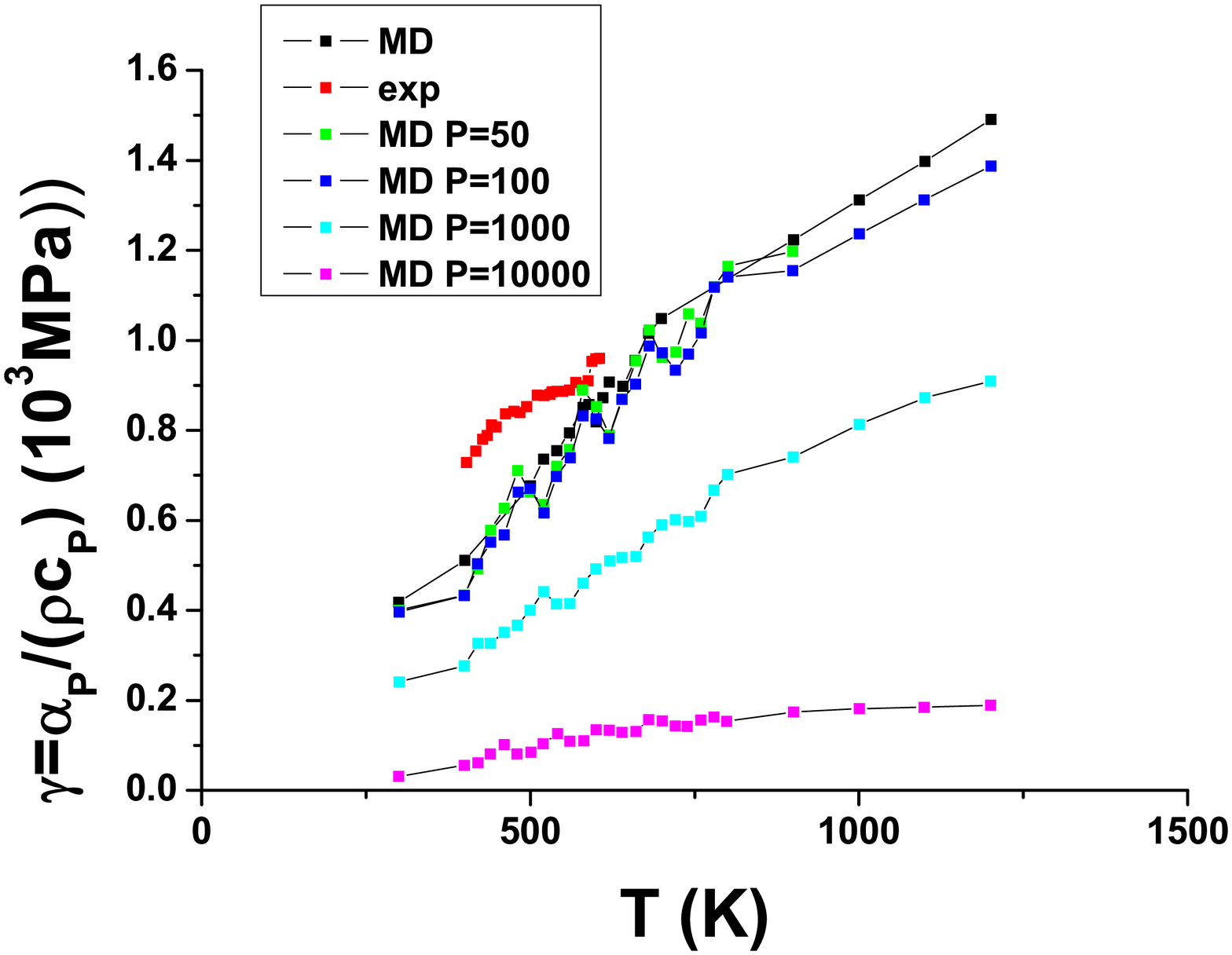}%

\caption{\label{g-all} Adiabatic thermal pressure coefficient
along several isobars.}
\end{figure}

\begin{figure}
\includegraphics[width=8cm]{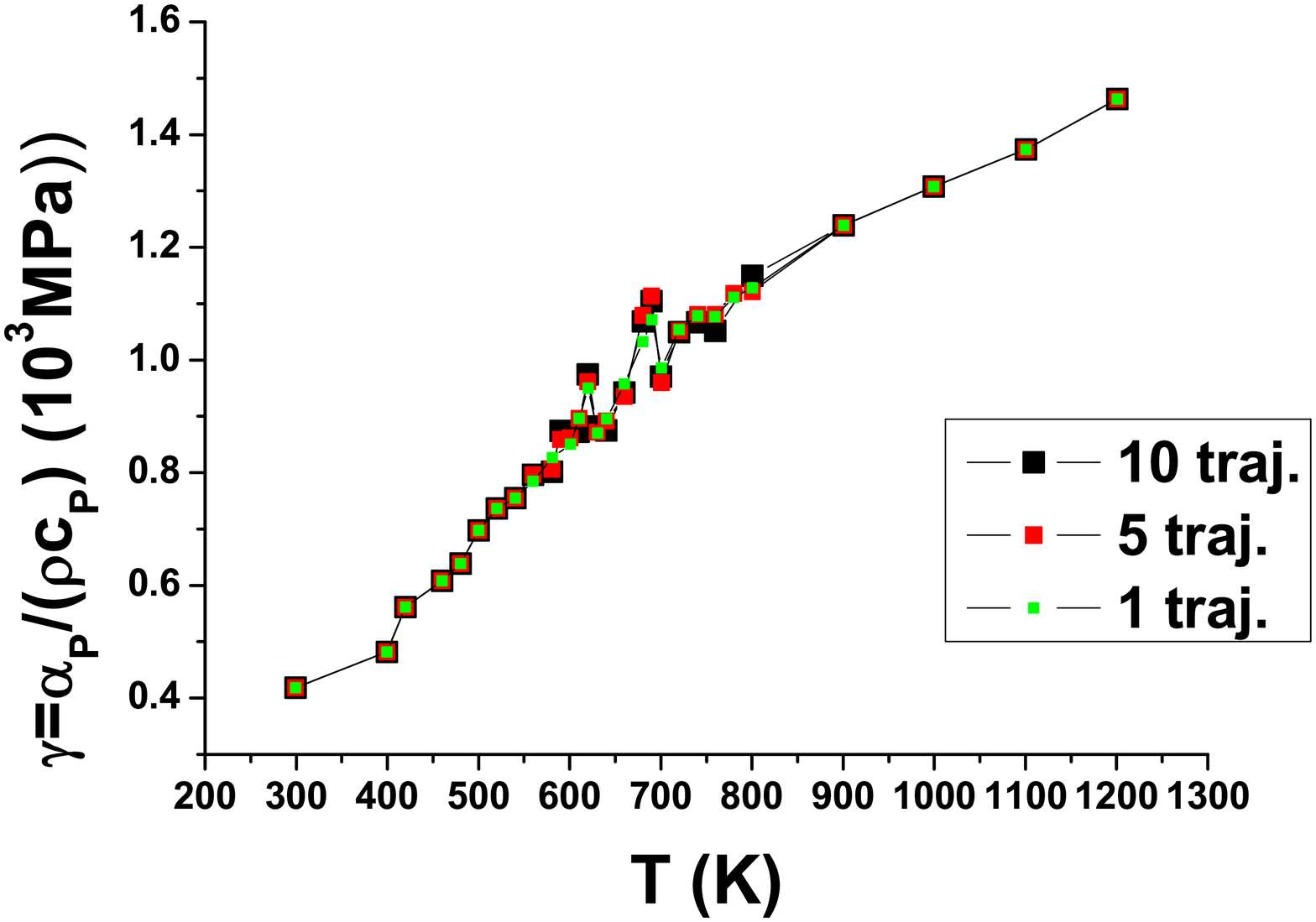}%

\caption{\label{g-10traj} Adiabatic thermal pressure coefficient
sampled $5$ and $10$ times more.}
\end{figure}

Finally, we study the behavior of the adiabatic thermal pressure
coefficient along several isobars in order to see if it
demonstrates some kind of jumps at other pressures. The results
are shown in Fig. \ref{g-all}. One can see that at all studied
pressures the curves are rather noisy which is, probably, related
with numerical differentiation of the results of the simulations
and does not allow to make any firm conclusion. Although we have
checked that our simulation converged, there is always systematic
noise in MD simulation which can be reduced by increasing the
averaging. In order to reduce the noise we perform additional
simulations. The simulation setup was the same as it was described
in the methods section. Some of the additional simulations were
performed starting from previously obtained configurations. Others
were started from different initial positions. In total 10
different simulations at each point were performed and the
averaging of the quantities of interest was done. The results are
shown in Fig. \ref{g-10traj}. One can see that the curve becomes
smoother, however, the kinks at $T=620$ and $690$ K preserved.
Therefore, these kinks can be real feature of the system under
investigation rather then numerical error, however, the physical
reason of these kinks is unclear.

In conclusion, we perform molecular dynamics study of probable
liquid-liquid phase transition in liquid cesium. The results of
the simulations do not show any traces of phase transition in such
quantities as equation of state, radial distribution function,
structure factor, angle distribution function and diffusion
coefficient. However, the adiabatic thermal pressure coefficient
demonstrates two small peaks which cannot be eliminated by
increasing the averaging. The origin of these peaks is not clear
and requires further studies.

\bigskip

We thank D.K. Belaschenko and R.M. Khusnutdinoff for help with EAM
potential construction and V.V. Brazhkin for useful discussions.
This work was carried out using computing resources of the federal
collective usage center "Complex for simulation and data
processing for mega-science facilities" at NRC "Kurchatov
Institute", http://ckp.nrcki.ru, and supercomputers at Joint
Supercomputer Center of the Russian Academy of Sciences (JSCC
RAS). The work was supported by the Russian Foundation of Basic Research (Grant No 18-02-00981).


\begin{thebibliography}{99}


\bibitem{tonkov} Tonkov E. Yu. and Ponyatovsky E. G. Phase
Transformations of Elements Under High Pressure. CRC PRESS; 2005.

\bibitem{exp-1990}  Tsuji K. et. al. Pressure-induced structural change of liquid cesium.  Journal of Non-Crystalline Solids 1990;
117-118: 72-75.


\bibitem{falconi} Falconi S, Lundegaard L F, Hejny C, and
McMahon M I. X-ray Diffraction Study of Liquid Cs up to 9.8 GPa.
Phys. Rev. Lett. 2005; 94: 125507.

\bibitem{hattori} Hattori T. Is there a pressure-induced discontinuous volume change in liquid Cs? Phys. Rev. B 2018; 97: 100101(R).




\bibitem{blag-97} Blagonravov L A, Kuznetsov S M, Alekseev V A,
Skovorod'ko S N. The behavior of variation of heat capacity of
liquid cesium in the region of anomalous behavior of structural
and thermodynamic parameters. High Temperature 1997; 35: 146–148.


\bibitem{blag-2000} Blagonravov L A et. al. Phase transition in liquid cesium near 590 K. J. of Non-Cryst.
Solids 2000; 277: 182.

\bibitem{blag-2008}  Blagonravov L A, Krylov A S, Mizotin M M , Skovorod'ko S N, and
Shpil'rain E E. The effect of structural phase transition on the
electrical resistivity of liquid cesium. High Temperature 2008;
46: 199–202.

\bibitem{blag-2012} Blagonravov L A, Soboleva A V, and Bogdanov N I. Thermophysical parameters of cesium in the
supercritical range. High Temperature 2012; 50: 52–55.

\bibitem{ghatee} Ghatee M H, Bahadori M. Characterizing phase transitions in liquid cesium by a soft-core and large-attraction equation of state.
Fluid Phase Equilibria 2005; 233: 151–156.

\bibitem{boat} New Kinds of Phase Transitions: Transformations in Disordered
Substances, Proceedings of the NATO Advanced Research Workshop,
Volga River, edited by Brazhkin V V, Buldyrev S V, Ryzhov V N, and
Stanley H E. Kluwer, Dordrecht; 2002.

\bibitem{genvdw} Fomin Yu D, Ryzhov V N, and Tareyeva E E. Generalized van der Waals theory of liquid-liquid phase
transitions. Phys. Rev. E 2006; 74: 041201.

\bibitem{skib} Skibinsky A, Buldyrev S V, Franzese G, Malescio G, and
Stanley H E. Liquid-liquid phase transitions for soft-core
attractive potentials. Phys. Rev. E 2004;  69: 061206.

\bibitem{skib1} Malescio G, Franzese G, Skibinsky A, Buldyrev S V, and
Stanley H E. Liquid-liquid phase transition for an attractive
isotropic potential with wide repulsive range. Phys. Rev. E 2005;
71: 061504.

\bibitem{rs-pre} Ryzhov V N and Stishov S M. Repulsive step potential: A model for a liquid-liquid phase transition. Phys. Rev. E 2003;
67: 010201(R) (2003).

\bibitem{rs-jetp} Ryzhov V N, Stishov S M. A liquid-liquid phase transition in the "collapsing" hard sphere system. JETP 2002; 95: 710.


\bibitem{eam-cs} Belashchenko D K. Structural transitions in liquid cesium. Russian Journal of Physical Chemistry
A 2014; 88: 1533.




\end{thebibliography}
\end{document}